\documentclass[showpacs,preprintnumbers,amsmath,amssymb]{revtex4}

\usepackage{graphicx}
\usepackage{dcolumn}
\usepackage{bm}
\usepackage{natbib}
\usepackage[dvips]{color}

\begin{document}


\title{Interoccurrence time statistics in the two-dimensional \\Burridge-Knopoff earthquake model}

\author{Tomohiro Hasumi}
 \email{t-hasumi.1981@toki.waseda.jp}

\affiliation{Department of Applied Physics, Advanced School of Science and Engineering, Waseda University, 3-4-1, Okubo, Shinjuku-ku, Tokyo 169-8555, Japan.}

\date{\today}

\begin{abstract}
We have numerically investigated statistical properties of the so-called interoccurrence time or the waiting time, i.e., the time interval between successive earthquakes, based on the two-dimensional (2-D) spring-block (Burridge-Knopoff) model, selecting the velocity-weakening property as the constitutive friction law. The statistical properties of frequency distribution and the cumulative distribution of the interoccurrence time are discussed by tuning the dynamical parameters, namely, a stiffness and frictional property of a fault. We optimize these model parameters to reproduce the interoccurrence time statistics in nature; the frequency and cumulative distribution can be described by the power law and Zipf-Mandelbrot type power law, respectively. In an optimal case, the $b$-value of the Gutenberg-Richter law and the ratio of wave propagation velocity are in agreement with those derived from real earthquakes. As the threshold of magnitude is increased, the interoccurrence time distribution tends to follow an exponential distribution. Hence it is suggested that a temporal sequence of earthquakes, aside from small-magnitude events, is a Poisson process, which is observed in nature. We found that the interoccurrence time statistics derived from the 2-D BK (original) model can efficiently reproduce that of real earthquakes, so that the model can be recognized as a realistic one in view of interoccurrence time statistics.
\end{abstract}

\maketitle

\section{\label{intro}introduction}
Earthquakes are complex phenomena, involving relative motion of faults. Many fundamental problems remain elusive, such as the source mechanism, a physical background, and so forth. To elucidate the source mechanism of earthquakes, it is essential to ascertain the physical background of the friction force acting on the surface of a fault. However, the unified friction law of a fault has not yet been established. Nowadays, some constitutive friction laws have been proposed based on laboratory experiments~\cite{Scholz:2002}, e.g., the velocity-weakening property~\cite{Scholz:1976, Dieterich:1978}, the rate and state dependent friction law~\cite{Dieterich:1979}, and the slip dependent constitutive law~\cite{Ohnaka:2003}. On the other hand, there are well-known and relevant empirical laws~\cite{Main:1996} for earthquakes. The most familiar one is the Gutenberg-Richter (GR) law~\cite{Gutenberg:1956}, which describes a relation between the seismic magnitude $M$ and its frequency $n$ as $n \propto  10^{-bM}$, where $b$ stands for the $b$-value and is similar to unity.
\par
The time interval between successive earthquakes is often called the interoccurrence time or waiting time. Very recently, statistical properties of the interoccurrence time have been studied~\cite{Ito:1995, Bak:2002, Wang:1998, Corral:2004, Abe:2005, Saichev:2006, Saichev:2007}. In Southern California, the probability density of the return time for a given $xy$-region of size $L$ follows a power law~\cite{Ito:1995}, which is governed by three factors~\cite{Bak:2002}: $b$-value of the GR law, the Omori law for aftershocks~\cite{Omori:1894}, and the fractal dimension of faults. Moreover, interoccurrence time distributions for large earthquakes exhibit both a gamma distribution and an exponential distribution~\cite{Wang:1998}. Corral showed after examining a wide variety of global earthquake catalogs that the probability density functions obey a generalized gamma distribution~\cite{Corral:2004}. Abe and Suzuki found that cumulative distributions are governed by the Zipf-Mandelbrot type power law~\cite{Mandelbrot:1983}, which is equivalent to the $q$-exponential distribution with $q>1$~\cite{Abe:2005} based on the nonextensive statistical mechanics proposed by Tsallis~\cite{Tsallis:1988}. Recently, Saichev and Sornette demonstrated that the distribution of interoccurrence time is derived from the known laws of the GR law and the Omori law~\cite{Saichev:2006, Saichev:2007}.\par 
In the 1980s, Bak~{\it et al.} proposed the concept of self-organized criticality (SOC)~\cite{Bak:1987, Bak:1988}, according to which a nonequilibrium open system gradually evolves into a critical state, where a distribution of physical quantity follows the power law. A crust is a nonequilibrium open system because energy is supplied by  plate motion and dissipated by an earthquake. Introducing a sand-pile model involving SOC and using the cellular automaton (CA) simulation method, they demonstrated an empirical power law which is related to the GR law~\cite{Bak:1989}. They suggested that earthquakes can be categorized into SOC phenomena. Inspired by their study, a number of SOC-based earthquake models have been proposed to reproduce statistical properties of earthquakes~\cite{Rundle:2003}. For example, a spring-block [Burridge-Knopoff (BK)] model~\cite{Burridge:1967} has been widely utilized. Based on the one-dimensional (1-D) BK model with nonlinear friction force, hereafter denoted by BK (original) model, Carlson and Langer showed the power law distribution like GR law, in a restricted magnitude region~\cite{Carlson:1989a,Carlson:1989b}. It has been shown that the power law distribution only concerns small creeping events~\cite{Carlson:1991b,Carlson:1994}. Carlson and Langer also discussed the interoccurrence time statistics for large earthquake events~\cite{Carlson:1991a}. Based on the 2-D BK (original) model extended by Carlson, the magnitude distribution~\cite{Carlson:1991d} and statistical properties of the shear stress drops~\cite{Kumagai:1999} were reported. This model calculation has also been performed by the CA simulation method, from here on referred to as the BK (CA) model, and the statistical properties of earthquakes, such as the magnitude distribution~\cite{Nakanishi:1990, Nakanishi:1991, Ito:1990, Otsuka:1972} and the interoccurrence time statistics~\cite{Brown:1991, Sanchez:2002} were discussed. Note that the 1-D BK (CA) model is not a good model for the Gutenberg-Richter law, see, for instance,~\cite{Schmittbuhl:1996}. The studies of the interoccurrence time statistics based on its modified versions were reported~\cite{Preston:2000, Christensen:1992, Hedges:2005,Hainzl:1999,Hainzl:2000}. The fundamental difference between the BK (original) model and the BK (CA) model is how to dissipate the friction energy; the friction force is governed by the constitutive law for the BK (original) model, and by the equipartition law for the BK (CA) model. The BK (original) model still has the great advantage of involving a constitutive friction law that well accounts for the laboratory experiment as well as the inertia.\par
The BK model in itself traces back to the late 1980s. We realize that the model simplifies the complex fault dynamics, so that some features of the fault, such as no radiation damping and long-range interactions, are inherently neglected. However, since the model can efficiently extract the statistical properties of earthquakes, such as the GR law and the Omori law~\cite{Nakanishi:1992}, it has been attracted much attention. Recent reports based on the BK (original) model have focused on the long range stress transfer~\cite{Xia:2005}, a fractal structure of faults~\cite{Hasumi:2006, Hasumi:2007a}, the statistical properties of the magnitude distribution and the interoccurrence time based on the rate and state dependent constitutive law~\cite{Omura:2007}, the correlation of seismicity~\cite{Mori:2005, Mori:2006, Kawamura:2006}, and the epicenter distance statistics~\cite{Hasumi:2007b}. However, the comprehensive application of the 2-D BK (original) model to the statistical properties of the interoccurrence time still remains to be done. According to the recent papers~\cite{Saichev:2006, Saichev:2007}, the interoccurrence time statistics derived from the modified versions of the BK model~\cite{Nakanishi:1992, Hedges:2005, Hainzl:1999, Hainzl:2000} which show the GR law and the Omori law reproduce these statistics in nature. On the other hand, in the case of the 2-D BK (original) model, whether the natural interoccurrence time statistics can be extracted or not is still contoroversial, because this model cannot produce aftershocks. In this study, we have attempted to work out how the interoccurrence time statistics are influenced by the variation of the major physical quantities, such as stiffness and frictional parameters of faults, and a threshold value of magnitude and by seismicity without aftershocks. Here, we report numerical investigations on the interoccurrence time statistics based on the 2-D BK (original) model by testing various dynamical parameters. These physical parameters are restricted or optimized so as to reproduce the statistics of actual earthquakes. We also compare the statistical properties derived from this model to those from the BK (CA) model. We show that the 2-D BK (original) model can successfully reproduce the interoccurrence time statics in nature.

\section{\label{model}Model}
\begin{figure*}
\begin{center}
\includegraphics{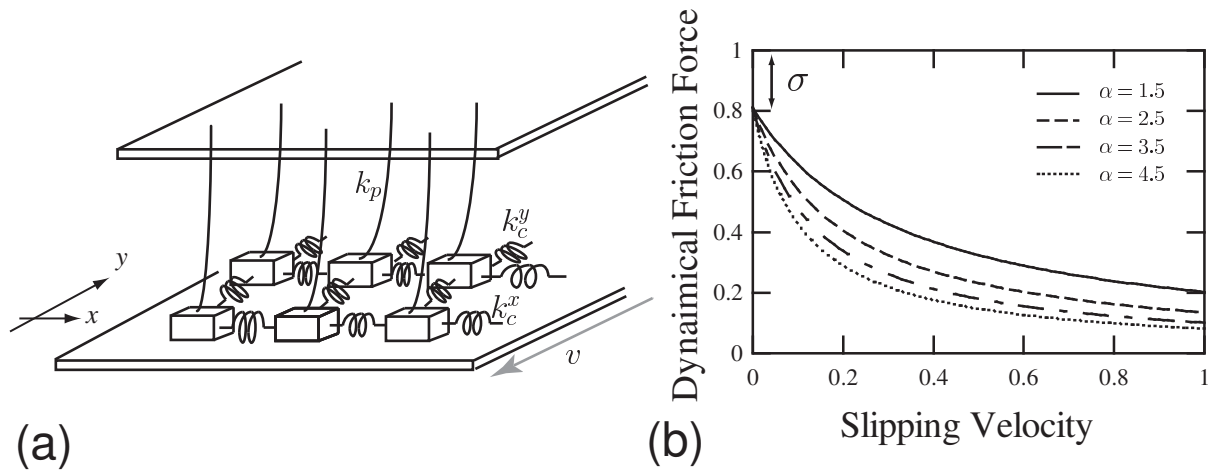}
\caption{(a) 2-D spring-block model. In our calculation, the system size is $(100\times 25)$ on the $(x,y)$ plane. $k_c^x$, $k_c^y$, and $k_p$ are spring constants. The friction force acts on the surface between the block and the bottom plate. (b) $\alpha$-dependence of the nonlinear dynamical friction function. $\sigma=0.01$ throughout the simulation.}
\label{BK_dim2}
\end{center}
\end{figure*}
In Fig.~1~(a), we schematically illustrate the 2-D spring-block model which we simulated in this study. The model is composed of blocks, two plates, two different kinds of coil springs, $k_c^x$ and $k_c^y$, and a leaf spring, $k_p$, corresponding respectively, to segments of a fault, geological plates, a compression stress, and a shear stress. The fundamental idea of the model is that earthquakes are caused by the stick-slip motion of the fault. It is noted that this model is virtually the same as Otsuka's model~\cite{Otsuka:1972} but is different from Carlson's~($k_c^x=k_c^y$)~\cite{Carlson:1991d}. We assume that the slip direction of blocks is restricted only to the $y$-direction. The equation of motion at site ($i,j$) can be written as
\begin{eqnarray}
m\frac{d^2 y_{i,j}}{dt^2} & &= k_c^x(y_{i+1,j}+y_{i-1,j}-2y_{i,j}) + k_c^y(y_{i,j-1}+y_{i,j+1}-2y_{i,j}) - k_p y_{i,j} -  F\left(v+\frac{dy_{i,j}}{dt}\right), 
\label{eqm_dim2}
\end{eqnarray}
where $m$ and $y_{i,j}$ are, respectively, mass and displacement of the block and $F$ is a dynamical force between the block and the bottom plate.\par  
Now we rewrite Eq.~(1) into a dimensionless form. The dimensionless dynamical friction force $\phi$ is obtained as $\phi (\dot{y}/v_1) = F(\dot{y})/F_0$, where $F_0$ and $v_1$ are the maximum friction force and the characteristic velocity, respectively. A dimensionless time $t'$ and displacement $U_{i,j}$ are defined by
\begin{eqnarray}
t' = \omega_p t = \sqrt{k_p/m}~t, \; \; U_{i,j} = \frac{y_{i,j}}{D_0}= \frac{y_{i,j}}{F_0/k_p}. \nonumber
\end{eqnarray}
Then the dimensionless form is 
\begin{eqnarray}
\frac{d^2 U_{i,j}}{dt'^2} & &= l_x^2(U_{i+1,j}+U_{i-1,j}-2U_{i,j}) +  l_y^2(U_{i,j-1}+U_{i,j+1}-2U_{i,j}) - U_{i,j} -  \phi \left[2\alpha \left(\nu+\frac{dU_{i,j}}{d t'}\right)\right],
\label{eqm_dim2_2}
\end{eqnarray}
and, 
\begin{eqnarray}
l_x^2 = \frac{k_c^x}{k_p}, \; \; l_y^2 = \frac{k_c^y}{k_p}, \; \; \nu = \frac{v}{D_0\omega_p} = \frac{v}{\hat{v}}, \; \; 2\alpha = \frac{D_0\omega_p}{v_1} = \frac{\hat{v}}{v_1}, \nonumber
\end{eqnarray}
where $\hat{v}$ is the slipping velocity, being on the order of 1 m/s in a real crust. In this work, we choose the friction force as the velocity-weakening constitutive law which states that as the slipping velocity increases, the dynamical friction force decreases. This constitutive law was observed in the accurate rock fracture experiment~\cite{Yoshida:1997}. Then, we use the dimensionless dynamical friction force $\phi$ given by 
\begin{eqnarray}
\phi (\dot{U}) = \left\{
\begin{array}{ll}
(-\infty, 1], & \dot{U}=0,\\
{\displaystyle \frac{(1-\sigma)}{\{1+2\alpha[\dot{U}/(1-\sigma)]\}}}, & \dot{U}>0.
\end{array}
\right.
\label{friction_function_2}
\end{eqnarray}
The formulation given in Eq.~(3) was introduced by Carlson~{\it et al.}~\cite{Carlson:1991b}~[see Fig.~1~(b)]. This can be regarded as an ideal friction function involving the stick-slip motion, so that it has often been adopted to 1-D and 2-D BK (original) models~\cite{Nakanishi:1992,Mori:2005,Mori:2006,Xia:2005,Hasumi:2006,Carlson:1991d,Kumagai:1999, Hasumi:2007a, Hasumi:2007b, Kawamura:2006}. To exclude a back slip, ($\dot{U}_{i,j}<0$), we treat $\phi$ as a tunable parameter when $\dot
{U}=0$.\par 
The model is governed by five parameters, $l_x, l_y, \alpha, \sigma$, and $\nu$. $l_x$ and $l_y$ are dimensionless stiffness parameters in the $x$- and $y$- direction, respectively. $\alpha$ means a decrement of the dynamical friction force with increasing slipping velocity. $\sigma$ is a stress gap between the normalized maximum friction force $(=1)$ and dynamical friction force $\phi(0)$. $\nu$ is a dimensionless loading velocity. $l_x$ and $l_y$ are related to Lame's constants, $\lambda$ and $\mu$, as~\cite{Yamashita:1976} 
\begin{eqnarray}
l_x^2 = \left(\frac{\Delta z}{\Delta x} \right)^2, \; \; l_y^2 = \frac{5\lambda +6\mu}{\lambda +2\mu}\left(\frac{\Delta z}{\Delta y} \right)^2, 
\label{Yamashita}
\end{eqnarray}
where $\Delta x,\Delta y$, and $\Delta z$ are infinitesimal lengths in the $x$-,$y$-, and $z$- directions, respectively. According to the experiments~\cite{Scholz:2002,Yoshida:1997}, $\alpha$ is positive and on the order of 10$^0$ (=1). We can treat $\nu$ as being equal to zero, when a slip event occurs because $\nu$, in nature $\sim 10^{-9}$, is far smaller than a slipping velocity of $\sim$ 1. This approximation ensures that no other event takes place during an ongoing event. Obviously, $\nu$ is related to loading time and interoccurrence time. We fixed $\nu=0.01$ because we will rescale it later. \par
In this paper, $\Delta x = \Delta y = \Delta z$ is used with assumptions of $\lambda=\mu$. We solved Eqs.~(2) and (3) by the fourth-order Runge-Kutta method under a free boundary condition. A small irregularity of block displacements is considered at an initial stage of the calculation. We use a $10^5$ order of earthquakelike events after some period when the initial randomness does not influence the statistical properties. Then the interoccurrence time statistics are systematically studied by changing $l_x$ and $l_y$, and by $\alpha$ increasing the threshold of magnitude $M_c$.

\section{Results and Discussions}
In this model, a slip of blocks is considered as an earthquake. An event starts when a block begins to slip in the direction toward $+y$ and ends when all blocks stop slipping. We define the interoccurrence time as the time interval between successive events. Accordingly, $n$th interoccurrence time can be described as $\tau_n=t'_n-t'_{n-1}$, where $t'_n$ and $t'_{n-1}$ are $n$th and $n-1$th earthquake occurrence time, respectively. 

\subsection{Friction parameter $\alpha$ dependence}
\begin{figure}
\includegraphics{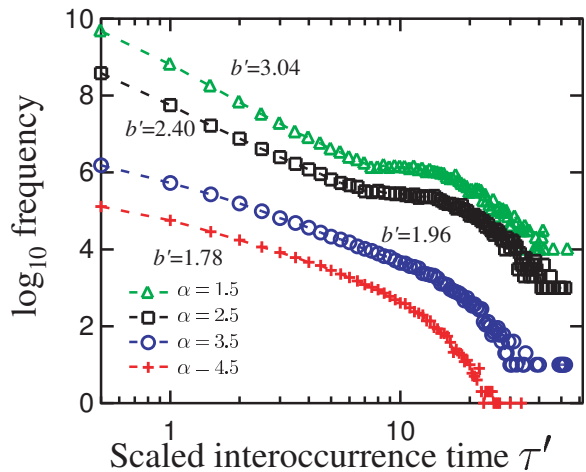}
\caption{The frequency distributions of the interoccurrence time for different $\alpha$~($\triangle:\alpha=1.5, \square: \alpha=2.5, \bigcirc:\alpha=3.5,$ and $+:\alpha=4.5$). $\bar{\tau}$ is the characteristic interoccurrence time, with fixing $l_x=1$ and $l_y=\sqrt{3}$. $b'$ is calculated from the slope of the dashed line. All plots except for in the case of $\alpha=4.5$ are shifted vertically for clarity.}
\label{power}
\end{figure}
In the first performance of our simulations, we study the $\alpha$- dependence of the interoccurrence time statistics. For that purpose, the stiffness parameters $l_x$ and $l_y$ are fixed ($l_x=1$ and $l_y=\sqrt{3}$), and the friction parameter $\alpha$ is changed from 1.5 to 4.5. As a result, the mean value of the interoccurrence time, $\bar{T}$, gradually increases as $\alpha$ is increased; namely, $\bar{T} \simeq  1.73~(\alpha=1.5),~2.65~(\alpha=2.5),~ 3.40~(\alpha=3.5)$, and $3.53~(\alpha=4.5)$. The frequency distributions of the interoccurrence time for different $\alpha$ are shown in Fig.~2 as a function of a scaled interoccurrence time, $\tau' = \tau/\bar{\tau}$, where $\bar{\tau}$ represents characteristic interoccurrence time and is set at 2.0 arbitrarily. We confirmed that in the case of $\alpha>4.5$, the statistical properties do not change qualitatively. The frequency distributions of any $\alpha$ exhibit a power law in the short-time region of $0.5 \lessapprox \tau' \lessapprox 7$. The power law exponent $b'$ gradually decreases as $\alpha$ is increased, for example, $b' \simeq 3.04~(\alpha=1.5)$, $b' \simeq 2.40~(\alpha=2.5)$, $b' \simeq 1.96~(\alpha=3.5)$, and $b' \simeq 1.78~(\alpha=4.5)$. As shown in Fig.~2, the distributions are categorized into three types; first, in the case of $\alpha=1.5$ and $2.5$, the frequency of a long interoccurrence time region of $\tau'> 10$ is enhanced more than expected by the power law decay, hereafter denoted by type A. Second, for $\alpha=4.5$, the frequency becomes less than predicted by the power law, which is referred to as type B. Finally, for the intermediate case of $\alpha=3.5$, the distribution most closely follows the power law, which is described as type C. It is suggested that the system undergoes a critical state when $\alpha \approx 3.5$ because the distribution shows the power law. We note that a critical power law has never been achieved qualitatively using the BK (original) model, whereas it has generally been obtained with the 2-D BK (CA) model~\cite{Hainzl:1999}. \par

\begin{figure*}
\includegraphics{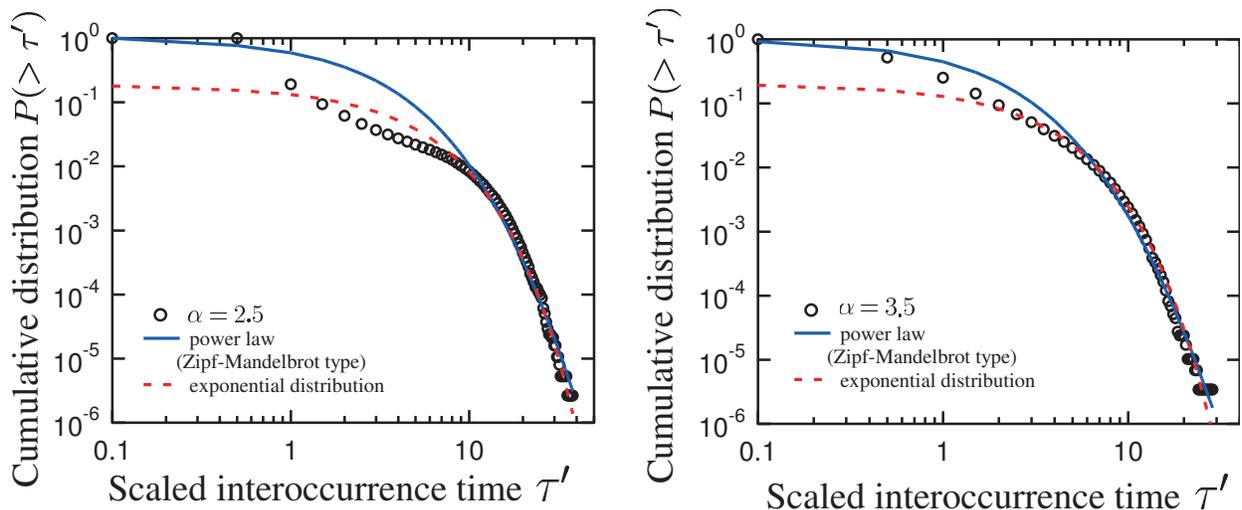}
\caption{The cumulative distributions of the interoccurrence time for $\alpha=2.5$ and 3.5 with fixing $l_x=1$ and $l_y=\sqrt{3}$. In this figure, the dotted, solid, and broken lines correspond, respectively, to the simulation data, the Zipf-Mandelbrot type power law defined in Eq.~(\ref{ZM}), and the exponential distribution.}
\label{TD}
\end{figure*}

Now, we discuss statistical properties of cumulative distributions of the interoccurrence time upon changing $\alpha$. According to~\cite{Abe:2005}, the cumulative distributions of the interoccurrence time, denoted by $P(>\tau')$, with earthquake data follow the Zipf-Mandelbrot type power law,
\begin{eqnarray}
P(>\tau') = \frac{1}{(1+\epsilon \tau')^\gamma}= e_q(-\tau'/\tau_0) =[ \left(1+(1-q)(-\tau'/\tau_0)\right)^{\frac{1}{1-q}}]_{+},
\label{ZM}
\end{eqnarray}
where $([a]_{+} \equiv  \textrm{max} [0,a])$ and $\epsilon,\gamma,q$, and $\tau_0$ are positive constants. Especially, $q,\tau_0$ and $e_q(x)$ are called an entropy index, a time-scale parameter, and the $q$-exponential function, respectively. $e_q(x)$ converges to an exponential function, $e^x$, as $q \to 1$. Figures.~3 (a) and 3~(b) are examples of $P(>\tau')$ for $\alpha=2.5$ and 3.5. In order to estimate an optimal $\alpha$, where the model yields or reproduces the interoccurrence time statistics in nature, we fit our numerical data to the Zipf-Mandelbrot power law given by Eq.~(5). As a result, these fitting parameters are estimated to be $q=1.10, \tau_0=2.26,$ and $\rho_z=0.986$ for $\alpha = 2.5$ (a) and $q=1.08, \tau_0=2.05,$ and $\rho_z=0.989$ for $\alpha=3.5$ (b) by the least-squares method. $\rho$ is a correlation coefficient between our simulation data and the fitting curve. For both $\alpha=2.5$ and 3.5, $q$ turns out to be closed to unity, so we introduce a different fitting model of the exponential distribution function described by $P(>\tau') = Ae^{-\tau'/{\tilde{\tau}}}$. The fitting parameters evaluated are $A=0.16, \tilde{\tau}=5.08$, and $\rho_e=0.988$ for $\alpha = 2.5$, and $A=0.18, \tilde{\tau}=3.97,$ and $\rho_e=0.980$ for $\alpha = 3.5$. The data demonstrate that in the case of $l_x=1$ and $l_y=\sqrt{3}$, $P(>\tau')$ can be best described by the Zipf-Mandelbrot type power law, simultaneously exhibiting quantitatively the nature of the real seismicity. Additionally, $P(>\tau')$ gives a good description of the Zipf-Mandelbrot type power law in the case of a small $b'$-value, judging from the small difference between our calculation and an ideal distribution, around $0.5 \lessapprox \tau' \lessapprox 3$. We note that the time-scale parameters, $\tau_0$ and $\tilde{\tau}$, are in particular sensitive to the variation of $\alpha$, compared to the other fitting parameters. 

\par

\subsection{Stiffness parameter $l_x$ and $l_y$ dependence}
In the second run of our simulations, the stiffness parameter-dependence of the interoccurrence time statistics is investigated, whereas $\alpha$ is fixed at 3.5, and $l_x$ and $l_y$ are systematically changed. We study the two different cases. One is an isotropic case, $l_x = l_y$, and the other is an anisotropic case, such as $l_x=1$ and $l_y =\sqrt{3}$. As a result, $\bar{T}$ depends on $l_x$ and $l_y$, for instance, $\bar{T} \simeq 3.94~(l_x= l_y =1)$, 2.24 $(l_x = l_y = \sqrt{3})$, 0.973 $(l_x = \sqrt{3}$ and $l_y = 3)$, and 2.08 $(l_x = \sqrt{2}$ and $l_y = \sqrt{5})$. In Fig.~4, we display the frequency distributions for different stiffness parameters. The $b'$-values also depend on the stiffness parameters, such as $b' \simeq 2.67~(l_x = l_y = \sqrt{3})$, $b' \simeq 3.09~(l_x = \sqrt{3}$ and $l_y = 3)$, $b' \simeq 2.65~(l_x = \sqrt{2}$ and $l_y = \sqrt{5})$, and $b' \simeq 1.94~(l_x = 1$ and $l_y = \sqrt{10})$. Thus we point out that among the systematically varied stiffness parameter settings, in almost all cases, the frequency distributions are categorized into type A, exhibiting a board peak in the long time region of $10 \lessapprox \tau' \lessapprox 30$. This implies that quasiperiodic events occur, which has been reported in the 1-D BK (original) model~\cite{Carlson:1991a, Mori:2005, Mori:2006} and in the 2-D BK (original) model~\cite{Carlson:1991d}, except for the case of $l_x=l_y=1$. \par
\begin{figure}
\includegraphics{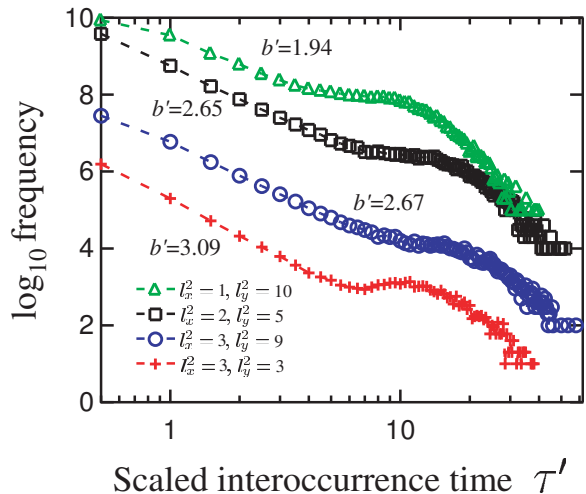}
\caption{The frequency distributions of the interoccurrence time for different $l_x$ and $l_y$~($\triangle: l_x=1$ and $l_y =\sqrt{10}, \square: l_x = \sqrt{2}$ and $l_y = \sqrt{5}, \bigcirc: l_x = l_y = \sqrt{3},$ and $+: l_x = \sqrt{3}$ and $l_y = 3$), with $\alpha=3.5$. All plots except for in the case of $l_x=1$ and $l_y =\sqrt{10}$ are shifted vertically for clarity. }
\label{frequency_l}
\end{figure}
As shown in Fig.~5, we discuss the cumulative distributions for different stiffness parameters, such as $l_x=l_y=\sqrt{3}$ (a) and $l_x=\sqrt{3}$ and $l_y=3$ (b). The Zipf-Mandelbrot type power law and the exponential distribution are used as distribution functions to fit our simulation results. Then, the Zipf-Mandelbrot power model yields $q=1.08$ and $\tau_0=2.67$ for (a), and $q=1.13$ and $\tau_0=1.23$ for (b), giving a correlation coefficient $\rho_z=0.954$ and 0.921, respectively. As for the exponential distribution, $A=0.08$ and $\tau_0=5.52$ for (a) and $A=0.017$ and $\tau_0=4.46$ for (b) are deduced together with $\rho_e=0.974$ and 0.962, respectively, for (a) and (b). \par
\begin{figure*}
\includegraphics{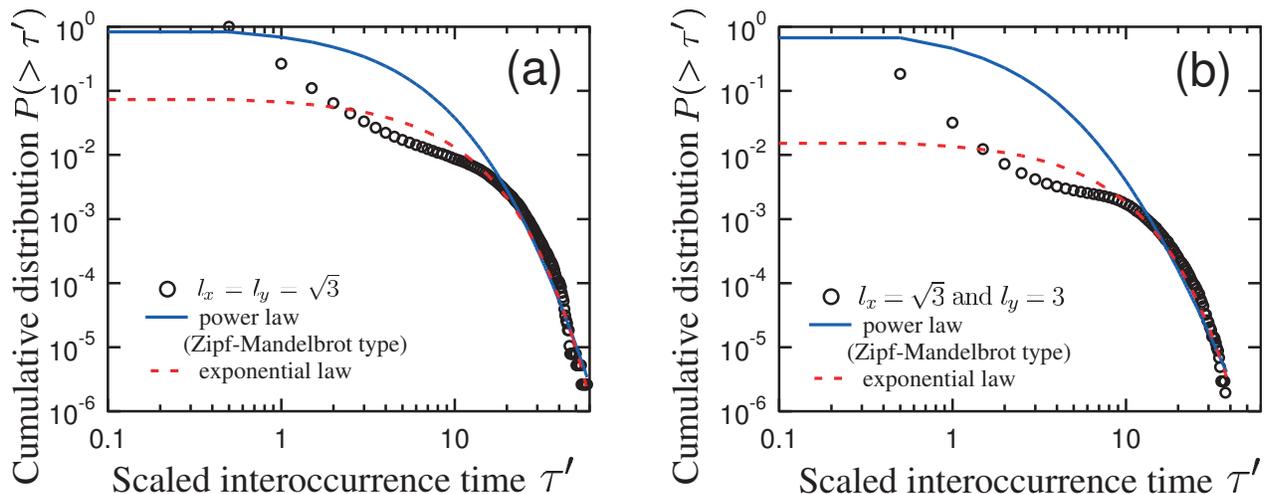}
\caption{The cumulative distributions of the interoccurrence time for $l_x = l_y = \sqrt{3}$ in (a) and $l_x = \sqrt{3}$ and $l_y = 3$ in (b), while $\alpha$ is fixed to be 3.5. In this figure, the dotted, solid, and broken lines correspond, respectively, to the simulation data, the Zipf-Mandelbrot type power law defined in Eq.~(\ref{ZM}), and the exponential distribution.}
\label{cumu_l}
\end{figure*}

\subsection{Threshold of magnitude $M_c$ dependence}
We study the magnitude-dependence of the interoccurrence time in order to discuss the interoccurrence time statistics for large magnitude earthquakes. In this model, we defined a seismic moment $M_0$ and a seismic magnitude $M$ as $M_0 = \sum_{i,j}^n \delta U_{i,j}$ and $M = (\log M_0)/1.5,$ where $\delta U_{i,j}$ is a total displacement at site $(i,j)$ during an event and $n$ is the number of slipping blocks~\cite{Ben-Menahem:1981}. One may find it unrealistic at first sight that the seismic magnitude $M$ can be negative. However, it is natural in the case where $n \lessapprox 10$, because the dimensionless displacement per block is less than unity by definition. Here, we examine the interoccurrence time statistics by altering $M_c$, using the simulation data for $l_x=1, l_y = \sqrt{3}$, and $\alpha=3.5$. It should be noted that the upper limit of $M_c$ is optimized so as to ensure sufficient data for us to evaluate the statistical properties.\par
\begin{figure*}
\includegraphics{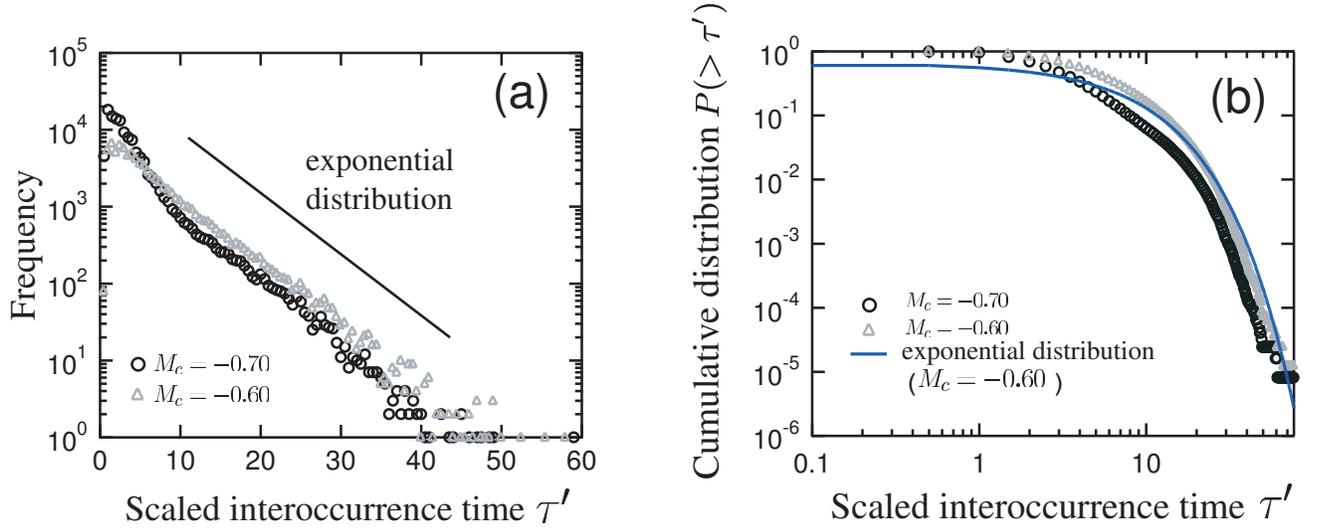}
\caption{The interoccurrence time statistics increasing the threshold of magnitude $M_c$, for $l_x=1,l_y=\sqrt{3}$, and $\alpha=3.5$ ($\bigcirc:M_c=-0.70,$ and $\triangle:M_c=-0.60$). The solid line corresponds to the exponential distribution. The frequency and cumulative distributions are in (a) and in (b), respectively.}
\label{threshold}
\end{figure*}
For the parameter setting given above, the seismic magnitude $M$ ranges from $-$0.86 to 1.5. In order to keep the statistical quantities accurate, we determined the upper limit of $M_c$ to be $-0.60$. Then, we selected $-$0.80, $-$0.75, $-$0.70, $-$0.65, and $-$0.60 as suitable $M_c$ values to test. The frequency distributions $M_c = -0.70$ and $-$0.60 are presented in Fig.~6~(a). The frequency of short interoccurrence times satisfying $0.5 \lessapprox \tau' \lessapprox 2$ gradually decreases when we increase $M_c$. This indicates that small magnitude events occur successively. It is found that the frequency distributions follow the exponential distribution for large $M_c$.  This implies that the cumulative distributions are also described by the exponential distribution. As shown in Fig.~6~(b), indeed the exponential law does govern the cumulative distributions for large $M_c$. Note that based on the 2-D BK (CA) model, the distribution for large $M_c$ is governed by the power law~\cite{Christensen:1992}, which differs from our results for the 2-D BK (original) model.\par
We can demonstrate that the interoccurrence time statistics depends on the threshold of magnitude $M_c$; the frequency and the cumulative distributions are definitely changed qualitatively and quantitatively as $M_c$ is increased. For large $M_c$, these distributions follow the exponential distributions. Moreover, it is found that a temporal sequence of events, except for small magnitude events, is a Poisson process, which is also reported using real earthquake data~\cite{Wang:1998}. We point out that the transition magnitude point can be estimated to be $-0.65$.

\subsection{Optimization of the model parameters and comparison with the real seismicity}
\begin{table*}
\caption{\label{table2}Summay of the interoccurrence time statistics based on the 2-D BK (original) model and on the real seismicity.}
\begin{ruledtabular}
\begin{tabular}{ccccccc}
$l_x$&$l_y$& $\alpha$&Frequency&Cumulative&$b$-value & P-wave/S-wave\\
 & & &distribution\footnotemark[1]&distribution\footnotemark[2]& &  \\
 
\hline
1 & $\sqrt{3}$ & 1.5 & Type A
& Power law  & $\sim 1.57$ & 1.7 \\
1 & $\sqrt{3}$ & 2.5 & Type A
& Exponential law  & $\sim 1.15$ & 1.7 \\
1 & $\sqrt{3}$ & 3.5 & Type C
& Power law &$\sim 1.00$ & 1.7 \\
1 & $\sqrt{3}$ & 4.5 & Type B
& Power law &$\sim 0.88$ & 1.7 \\
1 & 1 & 3.5 & Type B
& Power law &$\sim 1.11$ & 1.0 \\
$\sqrt{3}$ & $\sqrt{3}$ & 3.5 & Type A
& Exponential law &$\sim 1.54$ & 1.0 \\
$\sqrt{2}$ & $\sqrt{5}$ & 3.5 & Type A
& Exponential law &$\sim 1.31$ & 1.6 \\
$\sqrt{3}$ & 3 & 3.5 & Type A
& Exponential law & $\sim 1.37$ & 1.7 \\
\multicolumn{3}{c}{Real data}& Type C
& Power law~\cite{Abe:2005} & $\sim 1.0$ & 1.7\\ 
\end{tabular}
\end{ruledtabular}
\footnotetext[1]{As we mentioned before the frequency distributions are categorized as three types, namely, type A, type B, and type C.}
\footnotetext[2]{Two test distribution functions are introduced, namely the Zipf-Mandelbrot power law and the exponential distribution denoted here the power law and exponential law, respectively.}
\end{table*}

Now, we compare the interoccurrence time statistics obtained from our model with that observed in nature in order to confirm our model is realistic in view of extracting the interoccurrence time statistics of an actual earthquake. Here, the dynamical parameters are tuned so as to extract the statistical properties of earthquakes as clearly as possible. As displayed in Table~I, the optimal parameters are estimated to be $l_x=1, l_y = \sqrt{3}$, and $\alpha=3.5$. It is noted that the exponent $b'$ in nature is similar to unity because this value is related to the Omori law as $b' = 2 - (1/p)$~\cite{Utsu:1995}, where $p$ characterizes the Omori law. Our findings are in agreement with real data semiquantitatively because the $b'$-value is greater than 1. \par
\begin{figure*}
\includegraphics{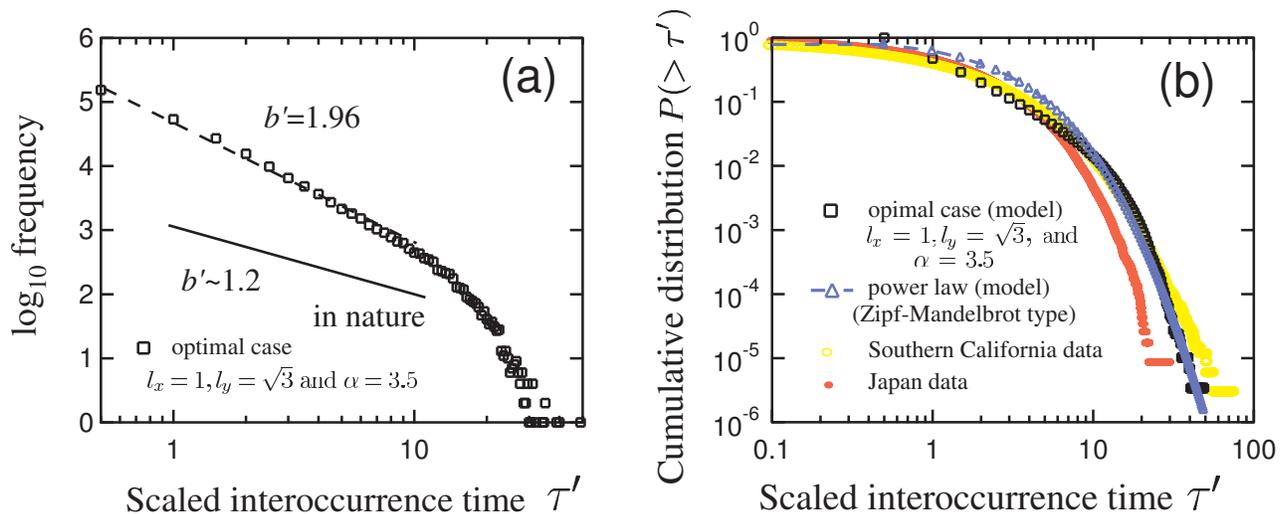}
\caption{The interoccurrence time statistics obtained from the model and real earthquake data. The frequency distribution and the cumulative distribution are shown in (a) and (b), respectively. In (b) the fitting parameter of the cumulative distribution yields $q=1.13$ and $\tau_0=1.72$, $q=1.05$ and $\tau_0=1.58$, and $q=1.08$ and $\tau_0=2.05$ from the Southern California data, and the Japan data, and our simulation, respectively.}
\label{real_data}
\end{figure*}
We display the cumulative distribution of interoccurrence time for Southern California and Japan overlapped in our simulation data in the case of the optimal parameters, namely, $l_x=1, l_y=\sqrt{3}$, and $\alpha=3.5$ in Fig.~7~(b). It is remarked that in order to compare the statistical properties, we used the data from~\cite{Abe:2005} and then we rescaled the interoccurrence time by $\bar{\tau}=1000$ s. It is found that the cumulative distributions also qualitatively satisfied those of real seismicity, the Zipf-Mandelbrot type power law. As can be seen from Figs. 3 (b) and Fig. 8 (b), there appears a small but distinguishable difference between our numerical data and the Zipf-Mandelbrot power law in the region of $0.4 \lessapprox \tau' \lessapprox 4$. When the $b'$-value gradually decreases, our data of $P(>\tau')$ become closed to the Zipf-Mandelbrot type power law. If the model reproduced the Omori law, corresponding to $b' \sim 1.2$, we could obtain realistic interoccurrence time. Therefore to obtain more realistic interoccuurence time statistics, we need to select a model which reproduces the Omori law as well as the GR law. \par
Finally, we discuss other remaining statistical properties, such as $b$-value of the GR law and the ratio of the seismic waves. We calculate the $b$-value from the slope of the power law magnitude distribution. The results are listed in Table.~I. As for the optimal parameters, $l_x=1, l_y=\sqrt{3}$, and $\alpha=3.5$, the $b$-value is similar to unity, which is consistent with the $b$-value in nature. As already described, $l_x$ and $l_y$ are defined, respectively, as $\sqrt{k_c^x/k_p}$ and $\sqrt{k_c^y/k_p}$. Here, we can rewrite them into $l_x=\sqrt{\frac{k_c^x/m}{k_p/m}}$ and $l_y=\sqrt{\frac{k_c^y/m}{k_p/m}}$. $l_x$ and $l_y$ correspond respectively to the secondary wave (S-wave) and the primary wave (P-wave) because in the $x$- and $y$- direction, the oscillation direction of blocks is respectively perpendicular to and parallel to the wave propagation direction. The velocities of the P-wave and S-wave in real crust are closed to 7 and 4~km/h, respectively, so that a real observation gives $l_y/l_x \sim 1.7$. For $l_x=1$ and $l_y=\sqrt{3}$, the ratio $l_y/l_x$ well coincides with that of the real earthquakes. Although for other cases, e.g., $l_x=\sqrt{2}$ and $l_y=\sqrt{5}$, and $l_x=\sqrt{3}$ and $l_y=3$, the ratio is again similar to 1.7, the correlation coefficient is found to be worse than that of $l_x=1$ and $l_y=\sqrt{3}$. Therefore we found that the $b$-value of the GR law and the ratio of seismic waves satisfy those of earthquakes in nature in the case of optimal parameters ($l_x=1, l_y=\sqrt{3}$, and $\alpha=3.5$) which derived from the interoccurrence time statistics.

\subsection{System size dependence}

\begin{figure*}
\includegraphics{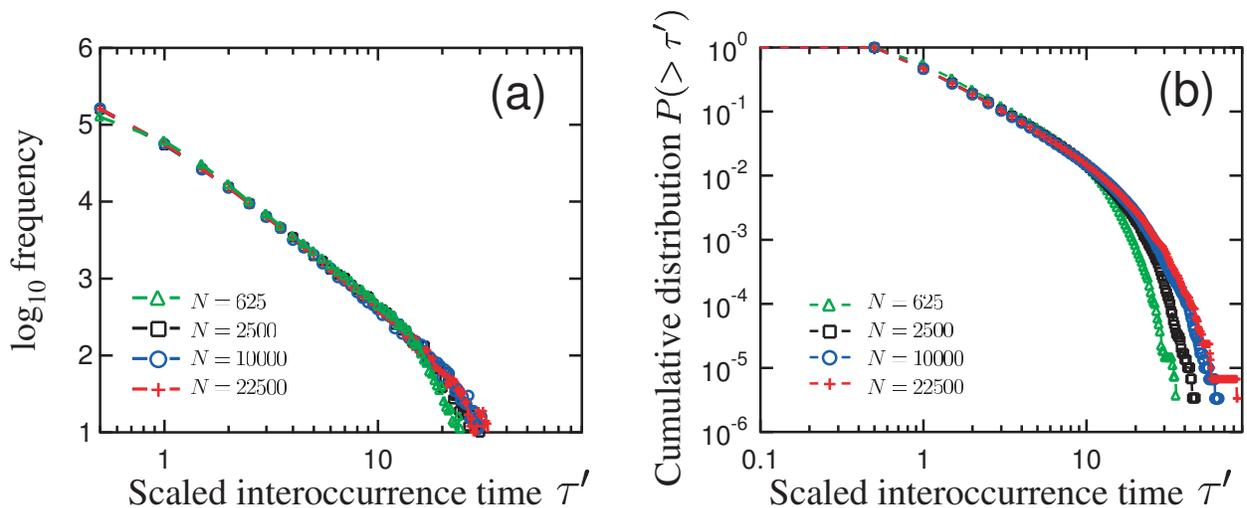}
\caption{Size dependence of the interoccurrence time statistics for the case of $l_x=1,l_y=\sqrt{3}$, and $\alpha=3.5$. We vary the system size $N$ from 625 to 22 500. ~($\triangle:N=625, \square: N=2500, \bigcirc:N=10\; 000,$ and $+:N=22\; 500$)}
\label{}
\end{figure*}

We study the size-dependence of the interoccurrence time statistics. For this purpose, the system size, $N$, is varied from $625~(25\times 25)$ to $22\; 500~(150\times 150)$, with fixing $l_x=1,l_y=\sqrt{3}$ and $\alpha=3.5$. In this case, the model reproduces a realistic $b$-value and the interoccurrence time statistics.\par
We display the interoccurrence time statistics for different system sizes, $N$, in Fig. 8. As clearly be seen from Fig. 8 (a), the frequency distribution shows the power law in the region of $1\ \lessapprox \tau' \lessapprox 20$ with an exponent, $b' \simeq 2.07$, 2.16, 2.18, and 2.16 for $N=625$, 2500, 10 000, and 22 500, respectively. This shows that the statistical property of the distribution, the power-law exponent $b'$, converges, for $N\gtrapprox 2500$. On the other hand, the maximum interoccurrence time gradually increases when $N$ is increased. It is shown in Fig. 8 (b) that the cumulative distributions do not depend on $N$, in the region of $0.1\lessapprox \tau'\lessapprox 10$, whereas it is found that the tail of the distributions becomes stretched for large $N$. By fitting the data to the Zipf-Mandelbrot power law given in eq.~(5), the optimal fitting parameters are obtained; $q=1.04$ and $\tau_0=2.34$ for $N=625$, $q=1.06$ and $\tau_0=2.55$ for $N=2500$, $q=1.11$ and $\tau_0=2.48$ for $N=10\; 000$, and $q=1.15$ and $\tau_0=2.10$ for $N=22\; 500$. The correlation coefficient, $\rho$, between our numerical result and the ideal curve is found to be 0.989, 0.986, 0.989, and 0.970 for $N=625$, 2500, 10 000, and 22 500, respectively. The result manifests that the statistical properties of $P(>\tau')$ remain vertually unchanged for $N\lessapprox 22\; 500$. Therefore, we conclude that the interoccurrence time statistics hold up when $2500 \lessapprox N \lessapprox 10\; 000$. It should be noted that in our numerical simulation, the system size, $N$, is set at 2500, where the size-dependence is negligible.

\section{Conclusion}
In light of our numerical investigations based on the 2-D BK (original) model, the interoccurrence time statistics depends on the dynamical parameters, stiffness, $l_x$ and $l_y$, and frictional properties of a fault, $\alpha$. Then we are able to restrict or optimize the dynamical parameters so as to reproduce of the interoccurrence time statistics in nature; the frequency distribution shows the power law and the cumulative distribution reveals the Zipf-Mandelbrot type power law. Additionally, we demonstrate that the interoccurrence time distributions follow the exponential distributions for large $M_c$, so that a temporal sequence of earthquakes, except for small-magnitude earthquakes, is reduced to a Poisson process. We have found that the 2-D BK (original) model with the optimal parameters $l_x=1, l_y=\sqrt{3}$, and $\alpha=3.5$ can be recognized as a realistic model for earthquakes in view of its fine reproduction of the major statistical properties of earthquakes, such as the interoccurrence time statistics, $b$-value of the GR law, and the ratio of the seismic waves. We conclude that the 2-D BK (original) model is sufficient to extract the realistic interoccurrence time statistics without employing the BK (CA) model that assumes an unrealistic friction law and no inertia. The present results underlie the advantages of the 2-D BK (original) model for its extensive application to other mechanical systems involving stick-slip behavior.
\begin{acknowledgments}
T.H. thanks Professor Y. Aizawa for encouragement and Dr. M. Kamogawa for useful comments, fruitful discussions, and manuscript improvements. This work was partly supported by the Japan Society for the Promotion of Science (JSPS), the Earthquake Research Institute cooperative research program at University of Tokyo, and a grant to the 21st Century COE Program, 'Holistic Research and Education Center for Physics of Self-organization Systems' at Waseda University from the Ministry of Education, Culture, Sports, Science, and Technology (MEXT), Japan.
\end{acknowledgments}

\appendix

\bibliographystyle{plain}

\end{document}